\documentclass[10pt,a4paper,onecolumn]{article}
\usepackage{marginnote}
\usepackage{graphicx}
\usepackage{xcolor}
\usepackage{authblk,etoolbox}
\usepackage{titlesec}
\usepackage{calc}
\usepackage{tikz}
\usepackage{hyperref}
\hypersetup{colorlinks,breaklinks=true,
            urlcolor=[rgb]{0.0, 0.5, 1.0},
            linkcolor=[rgb]{0.0, 0.5, 1.0}}
\usepackage{caption}
\usepackage{tcolorbox}
\usepackage{amssymb,amsmath}
\usepackage{ifxetex,ifluatex}
\usepackage{seqsplit}
\usepackage{xstring}

\usepackage{float}
\let\origfigure\figure
\let\endorigfigure\endfigure
\renewenvironment{figure}[1][2] {
    \expandafter\origfigure\expandafter[H]
} {
    \endorigfigure
}

\usepackage{fixltx2e} 
\usepackage[
  backend=biber,
]{biblatex}
\bibliography{paper.bib}

\let\textttOrig=\texttt
\def\texttt#1{\expandafter\textttOrig{\seqsplit{#1}}}
\renewcommand{\seqinsert}{\ifmmode
  \allowbreak
  \else\penalty6000\hspace{0pt plus 0.02em}\fi}

\makeatletter
\let\href@Orig=\href
\def\href@Urllike#1#2{\href@Orig{#1}{\begingroup
    \def\Url@String{#2}\Url@FormatString
    \endgroup}}
\def\href@Notdoi#1#2{\def\tempa{#1}\def\tempb{#2}%
  \ifx\tempa\tempb\relax\href@Urllike{#1}{#2}\else
  \href@Orig{#1}{#2}\fi}
\def\href#1#2{%
  \IfBeginWith{#1}{https://doi.org}%
  {\href@Urllike{#1}{#2}}{\href@Notdoi{#1}{#2}}}
\makeatother

\newlength{\cslhangindent}
\newlength{\csllabelwidth}
\newenvironment{CSLReferences}[3] 
 {
  \setlength{\parindent}{0pt}
  \ifodd #1 \everypar{\setlength{\hangindent}{\cslhangindent}}\ignorespaces\fi
  \ifnum #2 > 0
  \setlength{\parskip}{#2\baselineskip}
  \fi
 }%
 {}
\usepackage{calc}

\usepackage[top=3.5cm, bottom=3cm, right=2.54cm, left=2.54cm,
            headheight=2.2cm]{geometry}

\titleformat{\section}
  {\normalfont\sffamily\Large\bfseries}
  {}{0pt}{}
\titleformat{\subsection}
  {\normalfont\sffamily\large\bfseries}
  {}{0pt}{}
\titleformat{\subsubsection}
  {\normalfont\sffamily\bfseries}
  {}{0pt}{}
\titleformat*{\paragraph}
  {\sffamily\normalsize}

\usepackage{fancyhdr}
\makeatletter
\let\ps@plain\ps@fancy
\fancyheadoffset[L]{4.5cm}
\fancyfootoffset[L]{4.5cm}

\definecolor{linky}{rgb}{0.0, 0.5, 1.0}

\newtcolorbox{repobox}
   {colback=red, colframe=red!75!black,
     boxrule=0.5pt, arc=2pt, left=6pt, right=6pt, top=3pt, bottom=3pt}

\patchcmd{\@maketitle}{center}{flushleft}{}{}
\patchcmd{\@maketitle}{center}{flushleft}{}{}
\patchcmd{\@maketitle}{\LARGE}{\LARGE\sffamily}{}{}
\def\maketitle{{%
  
  \AB@maketitle}}
\makeatletter
\renewcommand\AB@affilsepx{ \protect\Affilfont}
\renewcommand\AB@affilnote[1]{{\bfseries #1}\hspace{3pt}}
\renewcommand{\affil}[2][]%
   {\newaffiltrue\let\AB@blk@and\AB@pand
      \if\relax#1\relax\def\AB@note{\AB@thenote}\else\def\AB@note{#1}%
        \setcounter{Maxaffil}{0}\fi
        \begingroup
        \let\href=\href@Orig
        \let\texttt=\textttOrig
        \let\protect\@unexpandable@protect
        \def\thanks{\protect\thanks}\def\footnote{\protect\footnote}%
        \@temptokena=\expandafter{\AB@authors}%
        {\def\\{\protect\\\protect\Affilfont}\xdef\AB@temp{#2}}%
         \xdef\AB@authors{\the\@temptokena\AB@las\AB@au@str
         \protect\\[\affilsep]\protect\Affilfont\AB@temp}%
         \gdef\AB@las{}\gdef\AB@au@str{}%
        {\def\\{, \ignorespaces}\xdef\AB@temp{#2}}%
        \@temptokena=\expandafter{\AB@affillist}%
        \xdef\AB@affillist{\the\@temptokena \AB@affilsep
          \AB@affilnote{\AB@note}\protect\Affilfont\AB@temp}%
      \endgroup
       \let\AB@affilsep\AB@affilsepx
}
\makeatother

\renewcommand\Affilfont{\sffamily\small\mdseries}
\ifnum 0\ifxetex 1\fi\ifluatex 1\fi=0 
  \usepackage[T1]{fontenc}
  \usepackage[utf8]{inputenc}

\else 
  \ifxetex
    \usepackage{mathspec}
    \usepackage{fontspec}

  \else
    \usepackage{fontspec}
  \fi
  \defaultfontfeatures{Ligatures=TeX,Scale=MatchLowercase}

\fi
\IfFileExists{upquote.sty}{\usepackage{upquote}}{}
\IfFileExists{microtype.sty}{%
\usepackage{microtype}
\UseMicrotypeSet[protrusion]{basicmath} 
}{}

\usepackage{hyperref}
\hypersetup{unicode=true,
            pdftitle={JAXbind: Bind any function to JAX},
            pdfborder={0 0 0},
            breaklinks=true}
\usepackage{color}
\usepackage{fancyvrb}

\DefineVerbatimEnvironment{Highlighting}{Verbatim}{commandchars=\\\{\}}
\newenvironment{Shaded}{}{}

\newcommand{\ControlFlowTok}[1]{\textcolor[rgb]{0.00,0.44,0.13}{\textbf{#1}}}

\newcommand{\DecValTok}[1]{\textcolor[rgb]{0.25,0.63,0.44}{#1}}

\newcommand{\FloatTok}[1]{\textcolor[rgb]{0.25,0.63,0.44}{#1}}

\newcommand{\ImportTok}[1]{#1}

\newcommand{\KeywordTok}[1]{\textcolor[rgb]{0.00,0.44,0.13}{\textbf{#1}}}
\newcommand{\NormalTok}[1]{#1}
\newcommand{\OperatorTok}[1]{\textcolor[rgb]{0.40,0.40,0.40}{#1}}

\let\addcontentslineOrig=\addcontentsline
\def\addcontentsline#1#2#3{\bgroup
  \let\texttt=\textttOrig\addcontentslineOrig{#1}{#2}{#3}\egroup}
\let\markbothOrig\markboth
\def\markboth#1#2{\bgroup
  \let\texttt=\textttOrig\markbothOrig{#1}{#2}\egroup}
\let\markrightOrig\markright
\def\markright#1{\bgroup
  \let\texttt=\textttOrig\markrightOrig{#1}\egroup}

\IfFileExists{parskip.sty}{%
\usepackage{parskip}
}{
\setlength{\parindent}{0pt}
\setlength{\parskip}{6pt plus 2pt minus 1pt}
}
\ifx\paragraph\undefined\else
\let\oldparagraph\paragraph
\renewcommand{\paragraph}[1]{\oldparagraph{#1}\mbox{}}
\fi
\ifx\subparagraph\undefined\else
\let\oldsubparagraph\subparagraph
\renewcommand{\subparagraph}[1]{\oldsubparagraph{#1}\mbox{}}
\fi

\title{\texttt{JAXbind}: Bind any function to JAX}

        \author[1, 2, 3*]{Jakob Roth}
          \author[1*]{Martin Reinecke}
          \author[1, 2, 4*]{Gordian Edenhofer}

      \affil[1]{Max Planck Institute for Astrophysics,
Karl-Schwarzschild-Str. 1, 85748 Garching, Germany}
      \affil[2]{Ludwig Maximilian University of Munich,
Geschwister-Scholl-Platz 1, 80539 Munich, Germany}
      \affil[3]{Technical University of Munich, Boltzmannstr. 3, 85748
Garching, Germany}
      \affil[4]{Department of Astrophysics, University of Vienna,
      T\"urkenschanzstr. 17, A-1180 Vienna, Austria}
      \affil[*]{These authors contributed equally.}
  \date{\vspace{-7ex}}

\begin{document}
\maketitle

\vspace{1em}

\hypertarget{summary}{%
\section{Summary}\label{summary}}

JAX is widely used in machine learning and scientific computing, the
latter of which often relies on existing high-performance code that we
would ideally like to incorporate into JAX. Reimplementing the existing
code in JAX is often impractical and the existing interface in JAX for
binding custom code either limits the user to a single Jacobian product
or requires deep knowledge of JAX and its C++ backend for general
Jacobian products. With \texttt{JAXbind} we drastically reduce the
effort required to bind custom functions implemented in other
programming languages with full support for Jacobian-vector products and
vector-Jacobian products to JAX. Specifically, \texttt{JAXbind} provides
an easy-to-use Python interface for defining custom, so-called JAX
primitives. Via \texttt{JAXbind}, any function callable from Python can
be exposed as a JAX primitive. \texttt{JAXbind} allows a user to
interface the JAX function transformation engine with custom derivatives
and batching rules, enabling all JAX transformations for the custom
primitive.

\hypertarget{statement-of-need}{%
\section{Statement of Need}\label{statement-of-need}}

The use of JAX (\protect\hyperlink{ref-Jax2018}{Bradbury et al., 2018})
is widespread in the natural sciences. Of particular interest is JAX's
powerful transformation system. It enables a user to retrieve arbitrary
derivatives of functions, batch computations, and just-in-time compile
code for additional performance. Its transformation system requires that
all components of the computation are written in JAX.

A plethora of high-performance code is not written in JAX and thus not
accessible from within JAX. Rewriting these codes is often infeasible
and/or inefficient. Ideally, we would like to mix existing
high-performance code with JAX code. However, connecting code to JAX
requires knowledge of the internals of JAX and its C++ backend.

In this paper, we present \texttt{JAXbind}, a package for bridging any
function to JAX without in-depth knowledge of JAX's transformation
system. The interface is accessible from Python without requiring any
development in C++. The package is able to register any function and its
partial derivatives and their transpose functions as a JAX native call,
a so-called primitive.

We believe \texttt{JAXbind} to be highly useful in scientific computing.
We intend to use this package to connect the Hartley transform and the
spherical harmonic transform from DUCC
(\protect\hyperlink{ref-ducc0}{Reinecke, 2024}) to the probabilistic
programming package NIFTy
(\protect\hyperlink{ref-Edenhofer2023NIFTyRE}{Edenhofer et al., 2024})
as well as the radio interferometry response from DUCC with the radio
astronomy package \texttt{resolve}
(\protect\hyperlink{ref-Resolve2024}{Arras et al., 2024}). Furthermore,
we intend to connect the non-uniform FFT from DUCC with JAX for
applications in strong-lensing astrophysics. We envision many further
applications within and outside of astrophysics.

The functionality of \texttt{JAXbind} extends the external callback
functionality in JAX. Currently, \texttt{JAXbind}, akin to the external
callback functions in JAX, briefly requires Python's global interpreter
lock (GIL) to call the user-specified Python function. In contrast to
JAX's external callback functions, \texttt{JAXbind} allows for both a
custom Jacobian-vector product and vector-Jacobian product. To the best
of our knowledge no other code currently exists for easily binding
generic functions and both of their Jacobian products to JAX, without
the need for C++ or LLVM. The package that comes the closest is
Enzyme-JAX (\protect\hyperlink{ref-Moses2024}{W. S. Moses \& Zinenko,
2024}), which allows one to bind arbitrary LLVM/MLIR, including C++,
with automatically-generated (\protect\hyperlink{ref-Moses2021}{W. S.
Moses et al., 2021}, \protect\hyperlink{ref-Moses2022}{2022};
\protect\hyperlink{ref-Moses2020}{W. Moses \& Churavy, 2020}) or
manually-defined derivatives to JAX.

PyTorch (\protect\hyperlink{ref-PyTorch2024}{Ansel et al., 2024}) and
TensorFlow (\protect\hyperlink{ref-tensorflow2015}{Abadi et al., 2015})
also provide interfaces for custom extensions. PyTorch has an
extensively documented Python interface\footnote{\url{https://pytorch.org/docs/stable/notes/extending.html}}
for wrapping custom Python functions as PyTorch functions. This
interface connects the custom function to PyTorch's automatic
differentiation engine, allowing for custom Jacobian and Jacobian
transposed applications, similar to what is possible with JAXbind.
Additionally, PyTorch allows a user to interface its C++ backend with
custom C++ or CUDA extensions\footnote{\url{https://pytorch.org/tutorials/advanced/cpp_extension.html}}.
JAXbind, in contrast, currently only supports functions executed on the
CPU, although the JAX built-in C++ interface also allows for custom GPU
kernels. TensorFlow includes a C++ interface\footnote{\url{https://www.tensorflow.org/guide/create_op}}
for custom functions that can be executed on the CPU or GPU. Custom
gradients can be added to these functions.

\hypertarget{automatic-differentiation-and-code-example}{%
\section{Automatic Differentiation and Code
Example}\label{automatic-differentiation-and-code-example}}

Automatic differentiation is a core feature of JAX and often one of the
main reasons for using it. Thus, it is essential that custom functions
registered with JAX support automatic differentiation. In the following,
we will outline which functions our package requires to enable automatic
differentiation via JAX. For simplicity, we assume that we want to
connect the nonlinear function \(f(x_1,x_2) = x_1x_2^2\) to JAX. The
\texttt{JAXbind} package expects the Python function for \(f\) to take
three positional arguments. The first argument, \texttt{out}, is a
\texttt{tuple} into which the results are written. The second argument
is also a \texttt{tuple} containing the input to the function, in our
case, \(x_1\) and \(x_2\). Via \texttt{kwargs\_dump}, any keyword
arguments given to the registered JAX primitive can be forwarded to
\(f\) in a serialized form.

\begin{Shaded}
\begin{Highlighting}[]
\ImportTok{import}\NormalTok{ jaxbind}

\KeywordTok{def}\NormalTok{ f(out, args, kwargs\_dump):}
\NormalTok{    kwargs }\OperatorTok{=}\NormalTok{ jaxbind.load\_kwargs(kwargs\_dump)}
\NormalTok{    x1, x2 }\OperatorTok{=}\NormalTok{ args}
\NormalTok{    out[}\DecValTok{0}\NormalTok{][()] }\OperatorTok{=}\NormalTok{ x1 }\OperatorTok{*}\NormalTok{ x2}\OperatorTok{**}\DecValTok{2}
\end{Highlighting}
\end{Shaded}

JAX's automatic differentiation engine can compute the Jacobian-vector
product \texttt{jvp} and vector-Jacobian product \texttt{vjp} of JAX
primitives. The Jacobian-vector product in JAX is a function applying
the Jacobian of \(f\) at a position \(x\) to a tangent vector. In
mathematical nomenclature this operation is called the pushforward of
\(f\) and can be denoted as \(\partial f(x): T_x X \mapsto T_{f(x)} Y\),
with \(T_x X\) and \(T_{f(x)} Y\) being the tangent spaces of \(X\) and
\(Y\) at the positions \(x\) and \(f(x)\). As the implementation of
\(f\) is not JAX native, JAX cannot automatically compute the
\texttt{jvp}. Instead, an implementation of the pushforward has to be
provided, which \texttt{JAXbind} will register as the \texttt{jvp} of
the JAX primitive of \(f\). For our example, this
Jacobian-vector-product function is given by
\(\partial f(x_1,x_2)(dx_1,dx_2) = x_2^2dx_1 + 2x_1x_2dx_2\).

\begin{Shaded}
\begin{Highlighting}[]
\KeywordTok{def}\NormalTok{ f\_jvp(out, args, kwargs\_dump):}
\NormalTok{    kwargs }\OperatorTok{=}\NormalTok{ jaxbind.load\_kwargs(kwargs\_dump)}
\NormalTok{    x1, x2, dx1, dx2 }\OperatorTok{=}\NormalTok{ args}
\NormalTok{    out[}\DecValTok{0}\NormalTok{][()] }\OperatorTok{=}\NormalTok{ x2}\OperatorTok{**}\DecValTok{2} \OperatorTok{*}\NormalTok{ dx1 }\OperatorTok{+} \DecValTok{2} \OperatorTok{*}\NormalTok{ x1 }\OperatorTok{*}\NormalTok{ x2 }\OperatorTok{*}\NormalTok{ dx2}
\end{Highlighting}
\end{Shaded}

The vector-Jacobian product \texttt{vjp} in JAX is the linear transpose
of the Jacobian-vector product. In mathematical nomenclature this is the
pullback \((\partial f(x))^{T}: T_{f(x)}Y \mapsto T_x X\) of \(f\).
Analogously to the \texttt{jvp}, the user has to implement this function
as JAX cannot automatically construct it. For our example function, the
vector-Jacobian product is
\((\partial f(x_1,x_2))^{T}(dy) = (x_2^2dy, 2x_1x_2dy)\).

\begin{Shaded}
\begin{Highlighting}[]
\KeywordTok{def}\NormalTok{ f\_vjp(out, args, kwargs\_dump):}
\NormalTok{    kwargs }\OperatorTok{=}\NormalTok{ jaxbind.load\_kwargs(kwargs\_dump)}
\NormalTok{    x1, x2, dy }\OperatorTok{=}\NormalTok{ args}
\NormalTok{    out[}\DecValTok{0}\NormalTok{][()] }\OperatorTok{=}\NormalTok{ x2}\OperatorTok{**}\DecValTok{2} \OperatorTok{*}\NormalTok{ dy}
\NormalTok{    out[}\DecValTok{1}\NormalTok{][()] }\OperatorTok{=} \DecValTok{2} \OperatorTok{*}\NormalTok{ x1 }\OperatorTok{*}\NormalTok{ x2 }\OperatorTok{*}\NormalTok{ dy}
\end{Highlighting}
\end{Shaded}

To just-in-time compile the function, JAX needs to abstractly evaluate
the code, i.e., it needs to be able to infer the shape and dtype of the
output of the function given only the shape and dtype of the input. We
have to provide these abstract evaluation functions returning the output
shape and dtype given an input shape and dtype for \(f\) as well as for
the \texttt{vjp} application. The output shape of the \texttt{jvp} is
identical to the output shape of \(f\) itself and does not need to be
specified again. The abstract evaluation functions take normal
positional and keyword arguments.

\begin{Shaded}
\begin{Highlighting}[]
\KeywordTok{def}\NormalTok{ f\_abstract(}\OperatorTok{*}\NormalTok{args, }\OperatorTok{**}\NormalTok{kwargs):}
    \ControlFlowTok{assert}\NormalTok{ args[}\DecValTok{0}\NormalTok{].shape }\OperatorTok{==}\NormalTok{ args[}\DecValTok{1}\NormalTok{].shape}
    \ControlFlowTok{return}\NormalTok{ ((args[}\DecValTok{0}\NormalTok{].shape, args[}\DecValTok{0}\NormalTok{].dtype),)}

\KeywordTok{def}\NormalTok{ f\_abstract\_T(}\OperatorTok{*}\NormalTok{args, }\OperatorTok{**}\NormalTok{kwargs):}
    \ControlFlowTok{return}\NormalTok{ (}
\NormalTok{        (args[}\DecValTok{0}\NormalTok{].shape, args[}\DecValTok{0}\NormalTok{].dtype),}
\NormalTok{        (args[}\DecValTok{0}\NormalTok{].shape, args[}\DecValTok{0}\NormalTok{].dtype),}
\NormalTok{    )}
\end{Highlighting}
\end{Shaded}

We have now defined all ingredients necessary to register a JAX
primitive for our function \(f\) using the \texttt{JAXbind} package.

\begin{Shaded}
\begin{Highlighting}[]
\NormalTok{f\_jax }\OperatorTok{=}\NormalTok{ jaxbind.get\_nonlinear\_call(}
\NormalTok{    f, (f\_jvp, f\_vjp), f\_abstract, f\_abstract\_T}
\NormalTok{)}
\end{Highlighting}
\end{Shaded}

\texttt{f\_jax} is a JAX primitive registered via the \texttt{JAXbind}
package supporting all JAX transformations. We can now compute the
\texttt{jvp} and \texttt{vjp} of the new JAX primitive and even
jit-compile and batch it.

\begin{Shaded}
\begin{Highlighting}[]
\ImportTok{import}\NormalTok{ jax}
\ImportTok{import}\NormalTok{ jax.numpy }\ImportTok{as}\NormalTok{ jnp}

\NormalTok{inp }\OperatorTok{=}\NormalTok{ (jnp.full((}\DecValTok{4}\NormalTok{,}\DecValTok{3}\NormalTok{), }\FloatTok{4.}\NormalTok{), jnp.full((}\DecValTok{4}\NormalTok{,}\DecValTok{3}\NormalTok{), }\FloatTok{2.}\NormalTok{))}
\NormalTok{tan }\OperatorTok{=}\NormalTok{ (jnp.full((}\DecValTok{4}\NormalTok{,}\DecValTok{3}\NormalTok{), }\FloatTok{1.}\NormalTok{), jnp.full((}\DecValTok{4}\NormalTok{,}\DecValTok{3}\NormalTok{), }\FloatTok{1.}\NormalTok{))}
\NormalTok{res, res\_tan }\OperatorTok{=}\NormalTok{ jax.jvp(f\_jax, inp, tan)}

\NormalTok{cotan }\OperatorTok{=}\NormalTok{ [jnp.full((}\DecValTok{4}\NormalTok{,}\DecValTok{3}\NormalTok{), }\FloatTok{6.}\NormalTok{)]}
\NormalTok{res, f\_vjp }\OperatorTok{=}\NormalTok{ jax.vjp(f\_jax, }\OperatorTok{*}\NormalTok{inp)}
\NormalTok{res\_cotan }\OperatorTok{=}\NormalTok{ f\_vjp(cotan)}

\NormalTok{f\_jax\_jit }\OperatorTok{=}\NormalTok{ jax.jit(f\_jax)}
\NormalTok{res }\OperatorTok{=}\NormalTok{ f\_jax\_jit(}\OperatorTok{*}\NormalTok{inp)}
\end{Highlighting}
\end{Shaded}

\hypertarget{higher-order-derivatives-and-linear-functions}{%
\section{Higher Order Derivatives and Linear
Functions}\label{higher-order-derivatives-and-linear-functions}}

JAX supports higher order derivatives and can differentiate a
\texttt{jvp} or \texttt{vjp} with respect to the position at which the
Jacobian was taken. Similar to first derivatives, JAX can not
automatically compute higher derivatives of a general function \(f\)
that is not natively implemented in JAX. Higher order derivatives would
again need to be provided by the user. For many algorithms, first
derivatives are sufficient, and higher order derivatives are often not
implemented by high-performance codes. Therefore, the current interface
of \texttt{JAXbind} is, for simplicity, restricted to first derivatives.
In the future, the interface could be easily expanded if specific use
cases require higher order derivatives.

In scientific computing, linear functions such as, e.g., spherical
harmonic transforms are widespread. If the function \(f\) is linear,
differentiation becomes trivial. Specifically for a linear function
\(f\), the pushforward or \texttt{jvp} of \(f\) is identical to \(f\)
itself and independent of the position at which it is computed.
Expressed in formulas, \(\partial f(x)(dx) = f(dx)\) if \(f\) is linear
in \(x\). Analogously, the pullback or \texttt{vjp} becomes independent
of the initial position and is given by the linear transpose of \(f\),
thus \((\partial f(x))^{T}(dy) = f^T(dy)\). Also, all higher order
derivatives can be expressed in terms of \(f\) and its transpose. To
make use of these simplifications, \texttt{JAXbind} provides a special
interface for linear functions, supporting higher order derivatives,
only requiring an implementation of the function and its transpose.

\hypertarget{platforms}{%
\section{Platforms}\label{platforms}}

Currently, \texttt{JAXbind} only supports primitives that act on CPU
memory. In the future, GPU support could be added, which should work
analogously to the CPU support in most respects. The automatic
differentiation in JAX is backend agnostic and would thus not require
any additional bindings to work on the GPU.

\hypertarget{acknowledgements}{%
\section{Acknowledgements}\label{acknowledgements}}

We would like to thank Dan Foreman-Mackey for his detailed guide
(https://dfm.io/posts/extending-jax/) on connecting C++ code to JAX.
Jakob Roth acknowledges financial support from the German Federal
Ministry of Education and Research (BMBF) under grant 05A23WO1
(Verbundprojekt D-MeerKAT III). Gordian Edenhofer acknowledges support
from the German Academic Scholarship Foundation in the form of a PhD
scholarship (``Promotionsstipendium der Studienstiftung des Deutschen
Volkes'').

\hypertarget{references}{%
\section*{References}\label{references}}
\addcontentsline{toc}{section}{References}

\hypertarget{refs}{}
\begin{CSLReferences}{1}{0}
\leavevmode\hypertarget{ref-tensorflow2015}{}%
Abadi, M., Agarwal, A., Barham, P., Brevdo, E., Chen, Z., Citro, C.,
Corrado, G. S., Davis, A., Dean, J., Devin, M., Ghemawat, S.,
Goodfellow, I., Harp, A., Irving, G., Isard, M., Jia, Y., Jozefowicz,
R., Kaiser, L., Kudlur, M., … Zheng, X. (2015). \emph{{TensorFlow}:
Large-scale machine learning on heterogeneous systems}.
\url{https://www.tensorflow.org/}

\leavevmode\hypertarget{ref-PyTorch2024}{}%
Ansel, J., Yang, E., He, H., Gimelshein, N., Jain, A., Voznesensky, M.,
Bao, B., Bell, P., Berard, D., Burovski, E., Chauhan, G., Chourdia, A.,
Constable, W., Desmaison, A., DeVito, Z., Ellison, E., Feng, W., Gong,
J., Gschwind, M., … Chintala, S. (2024, April). {PyTorch} 2: Faster
machine learning through dynamic {P}ython bytecode transformation and
graph compilation. \emph{29th ACM International Conference on
Architectural Support for Programming Languages and Operating Systems,
Volume 2 (ASPLOS ’24)}. \url{https://doi.org/10.1145/3620665.3640366}

\leavevmode\hypertarget{ref-Resolve2024}{}%
Arras, P., Roth, J., Ding, S., Reinecke, M., Fuchs, R., \& Johnson, V.
(2024). \emph{RESOLVE}. \url{https://gitlab.mpcdf.mpg.de/ift/resolve}

\leavevmode\hypertarget{ref-Jax2018}{}%
Bradbury, J., Frostig, R., Hawkins, P., Johnson, M. J., Leary, C.,
Maclaurin, D., Necula, G., Paszke, A., VanderPlas, J., Wanderman-Milne,
S., \& Zhang, Q. (2018). \emph{{JAX}: Composable transformations of
{P}ython+{N}um{P}y programs} (Version 0.3.13).
\url{http://github.com/google/jax}

\leavevmode\hypertarget{ref-Edenhofer2023NIFTyRE}{}%
Edenhofer, G., Frank, P., Roth, J., Leike, R. H., Guerdi, M.,
Scheel-Platz, L. I., Guardiani, M., Eberle, V., Westerkamp, M., \&
Enßlin, T. A. (2024). Re-envisioning numerical information field theory
({NIFTy.re}): A library for {G}aussian processes and variational
inference. \emph{Journal of Open Source Software}, \emph{9}(98), 6593.
\url{https://doi.org/10.21105/joss.06593}

\leavevmode\hypertarget{ref-Moses2021}{}%
Moses, W. S., Churavy, V., Paehler, L., Hückelheim, J., Narayanan, S. H.
K., Schanen, M., \& Doerfert, J. (2021). Reverse-mode automatic
differentiation and optimization of GPU kernels via enzyme.
\emph{Proceedings of the International Conference for High Performance
Computing, Networking, Storage and Analysis}.
\url{https://doi.org/10.1145/3458817.3476165}

\leavevmode\hypertarget{ref-Moses2022}{}%
Moses, W. S., Narayanan, S. H. K., Paehler, L., Churavy, V., Schanen,
M., Hückelheim, J., Doerfert, J., \& Hovland, P. (2022). Scalable
automatic differentiation of multiple parallel paradigms through
compiler augmentation. \emph{Proceedings of the International Conference
on High Performance Computing, Networking, Storage and Analysis}.
\url{https://doi.org/10.1109/SC41404.2022.00065}

\leavevmode\hypertarget{ref-Moses2024}{}%
Moses, W. S., \& Zinenko, O. (2024). \emph{{Enzyme-JAX}} (Version
0.0.6). \url{https://github.com/EnzymeAD/Enzyme-JAX}

\leavevmode\hypertarget{ref-Moses2020}{}%
Moses, W., \& Churavy, V. (2020). Instead of rewriting foreign code for
machine learning, automatically synthesize fast gradients. In H.
Larochelle, M. Ranzato, R. Hadsell, M. F. Balcan, \& H. Lin (Eds.),
\emph{Advances in neural information processing systems} (Vol. 33, pp.
12472–12485). Curran Associates, Inc.
\url{https://proceedings.neurips.cc/paper/2020/file/9332c513ef44b682e9347822c2e457ac-Paper.pdf}

\leavevmode\hypertarget{ref-ducc0}{}%
Reinecke, M. (2024). \emph{{DUCC}: Distinctly useful code collection}
(Version 0.33.0). \url{https://gitlab.mpcdf.mpg.de/mtr/ducc}

\end{CSLReferences}

\end{document}